\documentclass[amsmath,amssymb,12pt,superscriptaddress,nofootinbib]{revtex4}

\usepackage{graphicx}
\usepackage{dcolumn}
\usepackage{bm}
\usepackage{color}

\newcommand{\dd}{\mathrm{d}}
\newcommand{\rg}{r_{\rm g}}

\newcommand{\del}{\partial}
\definecolor{DarkBlue}{rgb}{0,0,0.7} 

\definecolor{DarkRed}{rgb}{0.65,0,0}

\begin{document}
\baselineskip5.5mm
\thispagestyle{empty}

{\baselineskip0pt
\leftline{\baselineskip14pt\sl\vbox to0pt{
               \hbox{\it Yukawa Institute for Theoretical Physics} 
              \hbox{\it Kyoto University}
               \vspace{1mm}
              \hbox{\it Department of Mathematics and Physics}
               \hbox{\it Osaka City  University}
               \vss}}
\rightline{\baselineskip16pt\rm\vbox to20pt{
            \hbox{YITP-12-29}
            {\hbox{OCU-PHYS-365}
            \hbox{AP-GR-97}
            }
\vss}}%
}

\author{Chul-Moon Yoo}\email{yoo@yukawa.kyoto-u.ac.jp}
\affiliation{
Yukawa Institute for Theoretical Physics, Kyoto University
Kyoto 606-8502, Japan
}

\author{Hiroyuki Abe}\email{abe@sci.osaka-cu.ac.jp}
\affiliation{ 
Department of Mathematics and Physics,
Graduate School of Science, Osaka City University,
3-3-138 Sugimoto, Sumiyoshi, Osaka 558-8585, Japan
}

\author{Yohsuke Takamori}\email{takamori@sci.osaka-cu.ac.jp}
\affiliation{
Osaka City University Advanced Mathematical 
Institute, 
3-3-138 Sugimoto, Sumiyoshi, Osaka 558-8585, Japan
}

\author{Ken-ichi Nakao}\email{knakao@sci.osaka-cu.ac.jp}
\affiliation{ 
Department of Mathematics and Physics,
Graduate School of Science, Osaka City University,
3-3-138 Sugimoto, Sumiyoshi, Osaka 558-8585, Japan
}

\vskip1cm
\title{Black Hole Universe \\ --- Construction and Analysis of Initial Data ---}

\begin{abstract}
We numerically construct an one-parameter family of 
initial data of an expanding inhomogeneous universe 
model
which is composed of regularly aligned 
black holes with an identical mass. 
They are initial data for vacuum solutions 
of
the Einstein equations. 
We call this universe model the ``black hole universe" 
and analyze the structure of these initial data. 
We study the relation between the mean expansion rate of the 3-space, 
which corresponds to  
the Hubble parameter, and the mass density of black holes. 
The result implies that the same relation as that of the 
Einstein-de Sitter universe is 
realized in the limit of the large separation between neighboring 
black holes. 
The applicability of the cosmological Newtonian $N$-body simulation 
to the dark matter composed of black holes is also discussed. 
The deviation of the spatial metric of the cosmological Newtonian 
$N$-body system from that of the black hole universe is found to be 
smaller than about 1\% in a region distant from the particles, 
if the separation length between neighboring particles is 20 times larger than 
their gravitational radius. 
By contrast, the deviation of the square of 
the Hubble parameter of
the cosmological Newtonian $N$-body system from that 
of the black hole universe is 
about 20\% for the same separation length. 
\end{abstract}

\maketitle
\pagebreak

\section{introduction}
The homogeneous and isotropic universe model 
has enjoyed great success in explaining the observational data. 
By contrast, as anyone well knows, 
our universe is not exactly homogeneous and 
includes
a lot of objects which serve as local nonlinear inhomogeneity. 
Usually, effects of local non-linear structures 
on the global property 
of the universe are
considered by intuitive way or using some approximate methods. 
One of the effective ways to test the validity of our intuition or the approximation 
is to construct and study an exact or almost exact 
solution of the field equations, which may not be 
so realistic but should be able to fully describe non-linear effects 
in an inhomogeneous universe model. 

One example of exact inhomogeneous solutions is 
the so-called Swiss-cheese universe
model
\cite{Einstein:1945id,Einstein:1946zz}:  
the dust in arbitrary number of non-overlapping spherical regions is  
removed in a model of the homogeneous and isotropic universe
filled with dust, 
and then 
each removed region is filled with a Schwarzschild black hole 
of the same mass as that of the removed dust. 
The remaining dust filled region, 
which is corresponding to ``cheese", is 
playing a role of the glue to connect Schwarzschild patches.
However, due to the existence of the cheese region, 
the Swiss-cheese model may be too special to 
see significant effects of local inhomogeneities on the global 
evolution of the universe. 
Hence, it is important 
to study a universe model in which black holes 
are uniformly distributed
without the cheese region. 
We call such an inhomogeneous universe model 
the ``black hole universe" 
in this paper.

About this issue, one innovative work has been done by 
Lindquist and Wheeler in 1957\cite{RevModPhys.29.432} 
and this work has been recently revisited in Refs.~\cite{Clifton:2009jw,Uzan:2010nw}. 
They divided 
a virtual 3-sphere
into $N$ cells ($N=$5, 8, 16, 24, 120 and 600) 
and put a black hole portion of the Schwarzschild spacetime 
on a spherical region centered in each cell.
Then they derived the equation of motion for this ``lattice universe" 
from junction conditions between the Schwarzschild cell and the 3-sphere. 
It is demonstrated that the maximal radius of the lattice universe 
asymptotes to that of the corresponding homogeneous and isotropic 
closed universe 
filled with dust 
in the limit of the large number of black holes. 
Here we should note that the lattice universe is not an exact solution 
and there are gaps between each Schwarzschild black holes
(see a figure 3 in Ref.~\cite{RevModPhys.29.432}).

Our purpose in this paper is to numerically construct initial data of 
the black hole universe. 
As a first step, we consider regularly aligned 
black holes with an identical mass. 
By its symmetry, no anisotropic relative velocities 
between neighboring black holes 
will appear, and this system is similar to a cold gas, i.e., dust. 
In order to obtain such initial data sets, we consider a black hole 
at the center of a cubic 
region
and impose the periodic boundary conditions 
on its faces. 
A recipe for the initial data of the black hole universe  
and numerical procedure are 
given in Sec.~\ref{sec:recipe}. 
The degree of
inhomogeneity of the black hole universe is 
demonstrated in Sec.~\ref{sec:inhomogeneity} by 
calculating the traceless part of the 3-dimensional Ricci curvature 
tensor of the initial hypersurface. 
The structure of the initial hypersurface is investigated 
in Sec.~\ref{sec:horizons} by searching for 
horizons.

One of the 
fascinating
issues of inhomogeneous universe models is 
the so-called averaging problem. 
Naively, we expect that a universe model
with local inhomogeneities, such as the black hole universe, 
can be globally described by a homogeneous universe 
model on average. 
However, 
the effect of local inhomogeneities 
on the global expansion definitely exists  
and the expansion history may 
be different from that of the 
homogeneous and 
isotropic universe\cite{1987CQGra...4.1697E,1988PhRvL..61.2175F,
Russ:1996km,Buchert:1999er}. 
This issue has been discussed a lot 
in past years(see reviews\cite{Ellis:2005uz,Buchert:2007ik,Rasanen:2011ki} 
and references therein), 
however there 
are few analyses which are
applicable to inhomogeneous models with 
highly nonlinear metric inhomogeneity.\footnote{
%
One of few exceptional examples was given in Ref.~\cite{Kai:2006ws}. 
They studied the volume expansion rate 
of a kind of the Swiss-cheese model and showed that the 
cosmic volume expansion can be accelerated by non-linear inhomogeneities.
%
While we were writing this paper, Ref.~\cite{Clifton:2012qh} 
appeared. In Ref.~\cite{Clifton:2012qh}, the authors 
analytically constructed $N$-body solutions of 
Einstein's constraint equations 
by considering regularly arranged distributions of 
discrete masses in topological 3-spheres. 
Significant differences between our present
work and Ref.~\cite{Clifton:2012qh} 
is the spatial topology and the existence of the cosmic volume expansion.
In our present case, the spatial topology is ${\bf T}^3$ with one point removed, 
and the expansion rate is finite 
while the initial date sets considered in Ref.~\cite{Clifton:2012qh} 
have a topology of ${\bf S}^3$ with $N$ points removed, 
and their expansion rates vanish, i.e., 
time symmetric. 
One of remarkable advantages of our work over 
Ref.~\cite{Clifton:2012qh} is that a dynamical simulation 
of expanding universe is possible starting from our initial data, 
while only contracting universe is possible with the initial data 
given in Ref.~\cite{Clifton:2012qh}. 
}
To solve this issue, 
we need 
to rely on 
numerical relativity. 
Our work may be the first step for 
the concrete analysis of 
the effects of non-linear inhomogeneities 
in expanding universes. 
Although we cannot address the 
real time evolution of 
the black hole universe yet, 
the one-parameter family of initial data sets 
can be regarded as a fictitious time evolution 
of the black hole universe. 
Using the 
initial data sets, 
we study the cosmic volume expansion rate of  
the black hole universe model in Sec.~\ref{sec:averaging}.

The cosmological $N$-body simulation is a powerful tool 
for studying the structure formation in the universe  
by dealing with the motion of point particles,   
based on the cosmological Newtonian approximation. 
Since the interaction between these particles is the only gravity, the 
cosmological $N$-body simulation follows the time evolution of the dark matter 
in the cosmological context. The black hole is a candidate of the ingredient 
for the dark matter, and it is believed that the cosmological 
$N$-body simulation is applicable also to the black-hole dark matter. 
But it is quite non-trivial issue whether the point particles in the 
cosmological $N$-body simulation can be simply identified with black holes. 
Hence, it is important to see in what situation 
the cosmological Newtonian $N$-body simulation 
is valid for the black hole universe. 
This issue is discussed in Sec.~\ref{sec:newton}. 
Sec.~\ref{sec:summary} is devoted to a summary. 

In this paper, we use the geometrized units 
in which the speed of light and Newton's gravitational constant 
are one, respectively.

\section{Construction of Initial Data for the Black Hole Universe}
\label{sec:recipe}

\subsection{Constraint equations}
In this paper, we are interested in 
the initial data of the vacuum Einstein equations. 
The initial data of the Einstein equations is equivalent to 
intrinsic and extrinsic geometries of a spacelike hypersurface, i.e., 
the intrinsic metric $\gamma_{ij}$ 
and the extrinsic curvature $K_{ij}$, 
which represents how the spacelike hypersurface is embedded into the 
4-dimensional spacetime. 
These are partially determined by the following 
four components of the Einstein equations: 
\begin{eqnarray}
&&\mathcal R+K^2-K_{ij}K^{ij}=0, 
\label{eq:hamicon2}
\\
&&D_j K^j_{~i}-D_iK=0, 
\label{eq:momcon2}
\end{eqnarray}
where $\mathcal R$ and $D_i$ are 
Ricci curvature scalar and the covariant derivative 
with respect to the intrinsic metric $\gamma_{ij}$, 
respectively, and $K=\gamma^{ij}K_{ij}$. 
Equation \eqref{eq:hamicon2} is called the Hamiltonian constraint, 
whereas Eq.~\eqref{eq:momcon2} is called the momentum constraint.

Following an established procedure 
(see, e.g., Ref.~\cite{Gourgoulhon:2007ue}), 
we adopt the Cartesian spatial coordinate system and
rewrite $\gamma_{ij}$ and $K_{ij}$ as 
\begin{eqnarray}
&&\gamma_{ij}=\Psi^4\tilde \gamma_{ij}, \\
&&K^{ij}=\Psi^{-10}\left[\tilde D^iX^j+\tilde D^jX^i-\frac{2}{3}\tilde \gamma^{ij}
\tilde D_k X^k
+\hat A^{ij}_{\rm TT}\right]
+\frac{1}{3}\Psi^{-4}\tilde \gamma^{ij}K, 
\label{decomposition}
\end{eqnarray}
where $\Psi:=({\rm det}\gamma_{ij})^{1\over12}$, $\tilde D_i$ is 
covariant derivative with respect to the 
conformal metric $\tilde \gamma_{ij}$, and 
$\hat A_{\rm TT}^{ij}$ satisfies 
the transverse and traceless conditions,
\begin{equation}
\tilde D_j\hat A^{ij}_{\rm TT}=0~,~~\tilde \gamma_{ij}\hat A_{\rm TT}^{ij} =0. 
\end{equation}

The conformal factor 
$\Psi$ is determined so that the constraint 
equations are satisfied. 
The conformal metric $\tilde{\gamma}_{ij}$ has not six 
but five independent components due to the constraint 
${\rm det}\tilde{\gamma}_{ij}=1$. 
The three of the five components of $\tilde{\gamma}_{ij}$ 
can be always eliminated by the spatial coordinate 
transformation, 
and hence 
there are two physically meaningful components 
which can be freely chosen.

In the decomposition \eqref{decomposition},
mutually independent six components of $K_{ij}$ are expressed by 
$X^i$, $\hat A^{ij}_{\rm TT}$ and $K$. 
The longitudinal traceless part composed of 
$X^i$ is determined so that 
the constraint equations are satisfied, whereas the trace part $K$ is  
related to the degree of freedom 
to choose the foliation of the spacetime by the family of 
spacelike hypersurfaces, or in other words, time slicing. 
By contrast, the transverse and 
traceless part $\hat A^{ij}_{\rm TT}$ 
has two independent components 
which can be freely chosen. 
These two components of $\hat A^{ij}_{\rm TT}$ and 
the physically meaningful two components of 
$\tilde{\gamma}_{ij}$ are usually regarded as 
physical degrees of freedom to 
set initial data for gravitational waves. 

In order to avoid the cosmic volume expansion caused 
by artificial gravitational radiation, 
we assume trivial form of the conformal metric and 
no 
transverse and traceless part of the 
extrinsic curvature
\begin{eqnarray}
\tilde \gamma_{ij}&=&\delta_{ij}, \label{conf-metric}\\
\hat A^{ij}_{\rm TT}&=&0, 
\end{eqnarray}
where $\delta_{ij}$ is Kronecker's delta. 
As usual, we denote the inverse of $\tilde{\gamma}_{ij}$ 
by $\tilde{\gamma}^{ij}$ which is also equal to Kronecker's delta $\delta^{ij}$.
Then, Eqs.\eqref{eq:hamicon2} and \eqref{eq:momcon2} are written as 
\begin{eqnarray}
&&\triangle\Psi+\frac{1}{8}(\tilde L X)_{ij}
(\tilde L X)^{ij}\Psi^{-7}
-\frac{1}{12}K^2\Psi^5=0, 
\label{eq:hamicon3}
\\
&&\triangle X^i+\frac{1}{3}\partial^i\partial_jX^j-\frac{2}{3}\Psi^6\partial^iK
=0, 
\label{eq:momcon3}
\end{eqnarray}
where $\triangle$ is the flat Laplacian, $\partial _i$ is the ordinary derivative, and 
\begin{equation}
(\tilde L X )^{ij}:=\del ^i X^j+\del ^j X^i-\frac{2}{3}\delta^{ij}\del_k X^k. 
\end{equation}
Here, note that $X_i:=\tilde{\gamma}_{ij}X^j=X^i$ and 
$\partial^i:=\tilde{\gamma}^{ij}\partial_j=\partial_i$, 
$(\tilde{L}X)_{ij}=\tilde{\gamma}_{ik}\tilde{\gamma}_{jl}(\tilde{L}X)^{kl}=(\tilde{L}X)^{ij}$, etc. 
We solve these equations by assuming an appropriate functional 
form of $K$ as shown below. 

\subsection{Boundary condition and the trace of the extrinsic curvature}

\begin{figure}[htbp]
\begin{center}
\includegraphics[scale=0.6]{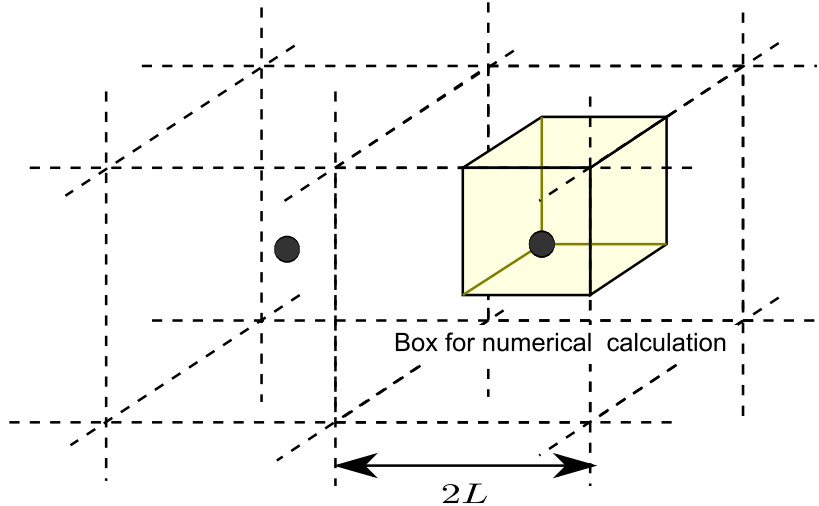}
\caption{The cubic region of our coordinates. 
}
\label{fig:cube}
\end{center}
\end{figure}

As mentioned above, 
we adopt the Cartesian coordinate system $\bm{x}=(x,y,z)$ and 
put a non-rotating black hole   
at the origin $\bm{x}=0$ denoted hereafter by $O$. 
The black hole is represented by 
a structure like the 
Einstein-Rosen bridge 
in our initial hypersurface. 
Thus the origin $O$ 
corresponds to the asymptotically flat spatial infinity and is 
often called the puncture.  
We focus on a cubic region 
$-L\leq x\leq +L$, $-L\leq y\leq +L$ and $-L\leq z\leq +L$ 
and 
call this region the domain $\mathcal D$. 
Since our aim is to construct the initial data of 
an expanding universe model 
with periodically aligned black holes, we impose the periodic boundary 
conditions; a point $\bm{x}=(-L,y,z)$ is identified with a point 
$\bm{x}=(+L,y,z)$, etc. 
Due to this boundary condition, the domain $\mathcal D$ is 
homeomorphic to the 3-torus ${\bf T}^3$. 
Since infinity is not included in the spacetime manifold,  
the initial hypersurface is ${\cal D}$ with $O$ removed, 
which is denoted by ${\cal D}-\{O\}$,  
and thus it is homeomorphic to ${\bf T}^3$ with one point removed.\footnote{
A similar configuration to our case was considered within the
Lema\^itre-Tolman family of exact models in Refs.~\cite{1987CQGra...4..635H,Plebanski:2006sd}}
The covering space of ${\cal D}-\{O\}$
represents a cosmological model
with regularly aligned black holes as shown in Fig.\ref{fig:cube}. 
Hereafter, we regard $\mathcal D$ as a cubic domain 
with boundary $\del \mathcal D$ in the covering space. 

Here, we again note that the trace part of the extrinsic curvature $K$ 
corresponds to the degree of freedom to choose the time slicing. 
In order to find 
the appropriate time slicing condition, first of all, we see 
the homogeneous and isotropic universe model. 
In this case, 
the expansion rate 
$H$ which is called the Hubble parameter 
is related to the extrinsic curvature by
\begin{equation}
H=-\frac{1}{3}K. 
\end{equation}
The above relation implies that $K$ of the expanding black hole universe
model must be negative 
at least around the boundary of the cubic domain $\mathcal D$. 
By contrast, 
the maximal slicing condition $K=0$ is appropriate for the foliation of 
the domain in the neighborhood of the asymptotically flat 
spatial infinity, and hence $K$ should vanish in the vicinity of $O$ 
(see Appendix\ref{sec:cmc}). 

\begin{figure}[htbp]
\begin{center}
\includegraphics[scale=1.2]{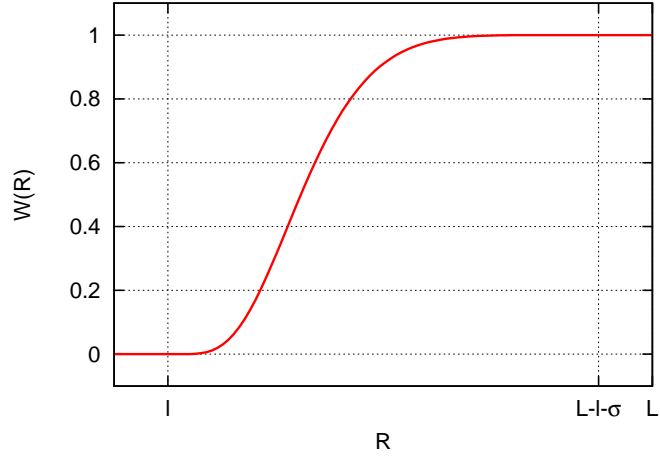}
\caption{The functional form of $W(R)$.
%
%
}
\label{fig:funcW}
\end{center}
\end{figure}

Taking the above discussions into account, 
we assume 
\begin{equation}
K(\bm x)=-3H_{\rm eff}W(R), 
\end{equation}
where $H_{\rm eff}$ is a positive constant 
which corresponds to the effective Hubble parameter,
$R:=|\bm x|$, and 
\begin{equation}
W(R)=
\left\{
\begin{array}{ll}
0&{\rm for}~0\leq R \leq \ell \\
\sigma^{-36}[(R-\sigma-\ell)^6-\sigma^6]^6&{\rm for}~\ell\leq R \leq \ell 
+ \sigma\\
1&{\rm for}~\ell+\sigma \leq R \\
\end{array}\right.,
\end{equation}
$\ell$ and $\sigma$ being constants 
which satisfy $\ell<\sigma<L$ (see Fig.\ref{fig:funcW}). 

\subsection{Extraction of the singularity at the center}

Since $K$ vanishes in the vicinity of the origin $O$, 
$X^i$ and $\Psi$ should behave as those of the Schwarzschild spacetime 
with the static isotropic coordinate system, 
\begin{eqnarray}
&&X^i\simeq 0, \\
&&\Psi\simeq\Psi_{\rm c}+\frac{M}{2R}, \label{Psi-def}
\end{eqnarray}
where $\Psi_{\rm c}$ and $M$ are positive constants. 
Since $\Psi$ itself is singular at $O$, 
we cannot handle $\Psi$ numerically. Thus, instead of $\Psi$, we 
solve the constraint equations for 
the following new variable $\psi$:
\begin{equation}
\psi(\bm x):=\Psi(\bm x)-\frac{M}{2R}\left[1-W(R)\right].  \label{psi-def}
\end{equation}
Thanks to the second term 
proportional to $[1-W(R)]$ in the right hand side of the above equation, 
$\psi$ is regular at $O$ and 
satisfies the periodic boundary conditions. 

The mass of a black hole is given by the ADM mass which is defined by the 
surface integral over the spacelike infinity 
at $O$. 
To see the ADM mass explicitly, we introduce a new radial coordinate 
\begin{equation}
\tilde{R}=\frac{M^2}{4R}.
\end{equation}
Then, by using a spherical polar coordinate system, the asymptotic form of the 
infinitesimal line element for $R\rightarrow 0$, or equivalently, $\tilde{R}\rightarrow\infty$, 
becomes 
\begin{eqnarray}
\dd l^2&\simeq&\left(\Psi_{\rm c}+\frac{M}{2R}\right)^4\left[\dd R^2+R^2(\dd\theta^2+\sin^2\theta \dd\phi^2)\right] \label{isotropic}\\
&=&\left(1+\frac{\Psi_{\rm c}M}{2\tilde{R}}\right)^4\left[\dd\tilde{R}^2+\tilde{R}^2(\dd\theta^2+\sin^2\theta \dd\phi^2)\right]. 
\label{asymptotic}
\end{eqnarray}
It is seen from the last equality of Eq.~\eqref{asymptotic} that 
the mass of a black hole is given by $\Psi_{\rm c} M$. 
Here note that 
there is a freedom of constant scaling of coordinates $\bm{x}\rightarrow C\bm{x}$. 
Using 
this freedom, we impose $\Psi_{\rm c}=1$, and thus 
the mass of a black hole is equal to $M$.

\subsection{Hubble equation as an integrability condition}

Integrating Eq.\eqref{eq:hamicon3}   
over the physical domain ${\cal D}-\{O\}$, 
we obtain the following equation: 
\begin{equation}
2\pi M
+\frac{1}{8}\int_{{\cal D}-\{O\}}  (\tilde L X)_{ij}(\tilde L X)^{ij}\Psi^{-7}\dd x^3
-\frac{3}{4}H_{\rm eff}^2 V=0,
\label{eq:effHub1}
\end{equation}
where, by noting that the origin $O$ 
can be regarded as 
the only boundary of ${\cal D}-\{O\}$
with the periodic boundary condition,  
the integral of $\Delta\Psi$ is rewritten as  
\begin{equation}
\int_{{\cal D}-\{O\}}\Delta \Psi \dd^3x
=-\lim_{R\rightarrow0}\int_0^{2\pi}\int_0^\pi \frac{\partial\Psi}{\partial R}
R^2\sin\theta \dd\theta \dd\phi 
=2\pi M, 
\end{equation}
and we have defined $V$ by
\begin{equation}
V:=\int_{{\cal D}-\{O\}}W^2\Psi^5\dd^3x. \label{V-def}
\end{equation}
By rewriting Eq.~\eqref{eq:effHub1}, we have the effective Hubble equation as
\begin{equation}
H_{\rm eff}^2=\frac{8\pi}{3}\left(\rho_{\rm BH}+\rho_{\rm K}\right),
\label{eq:effHub2}
\end{equation}
where $\rho_{\rm BH}$ and $\rho_{\rm K}$ are defined by
\begin{eqnarray}
\rho_{\rm BH}&:=&\frac{M}{V}, \label{eq:rhoBH}\\
\rho_{\rm K}&:=&\frac{1}{16\pi V}
\int_{{\cal D}-\{O\}} (\tilde L X)_{ij}(\tilde L X)^{ij}\Psi^{-7}\dd^3x. \label{eq:rhoK}
\end{eqnarray}
Since $V$ may be regarded as the effective volume of the expanding region, 
$\rho_{\rm BH}$ and $\rho_{\rm K}$ may be regarded as the mass density 
of black holes and the kinetic energy density of the spacetime, respectively. 
The effective Hubble equation gives a relation between 
two constants, the mass of the black hole $M$ and the effective 
Hubble parameter $H_{\rm eff}$, and, at the same time, 
it is an integrability condition of the constraint equations. 
How to guarantee this relation will be described in Sec.~\ref{sec:numeste}. 

%
\subsection{Momentum constraints}
In this subsection, we rewrite 
the momentum constraints \eqref{eq:momcon3} 
into the  
numerically solvable forms. 
First, we define $Z$ by
\begin{equation}
Z:=\del_iX^i. 
\end{equation}
Then, by taking the divergence of Eq.~\eqref{eq:momcon3}, we obtain 
\begin{eqnarray}
\triangle Z&=&\frac{1}{2}\del_i(\Psi^6\del^iK). 
\label{eq:forZ}
\end{eqnarray}
Eq.~\eqref{eq:momcon3} is rewritten as 
\begin{equation}
\triangle X^i=\frac{1}{3}\del^iZ+\frac{2}{3}\Psi^6\del^iK. 
\label{eq:forXi}
\end{equation}
The system we consider 
is unchanged if it rotates $2\pi/3$ radians around the line $x=y=z$. 
By virtue of this discrete symmetry, 
it is enough to solve Eq.~\eqref{eq:forXi} for only one component of $X^i$,  
since the other
two components can be immediately given by this symmetry. 

The boundary condition for $X^x$ is given as follows, 
\begin{eqnarray}
X^x&=&0~~{\rm on}~~x=0~~{\rm and}~~ x=L ~ , \label{B-1}\\
\del_yX^x&=&0~~{\rm on}~~y=0~~{\rm and}~~ y=L ~,\\
\del_zX^x&=&0~~{\rm on}~~z=0~~{\rm and}~~ z=L ~. \label{B-3} 
\end{eqnarray}
The first condition is the Dirichlet type, 
and the second and third ones are Neumann type boundary conditions.
These boundary conditions lead to 
\begin{equation}
\int_{{\cal D}-\{O\}}Z\dd x^3=
\int_{{\cal D}-\{O\}}\del_iX^i \dd x^3=0. \label{consistency}
\end{equation}
The above equation is a consistency condition that the solutions should satisfy. 

It should be noted that the integrals of the source terms of the Poisson equations 
\eqref{eq:forZ} and \eqref{eq:forXi} over ${\cal D}-\{O\}$ 
should vanish by the consistency 
with the periodic boundary conditions and the boundary condition at the origin $O$. 
We can see that these conditions are automatically satisfied. 
The integral of the source term of Eq.~\eqref{eq:forZ} is equivalent to the surface integral 
over the spatial infinity at $O$, 
whereas $K$ vanishes in the neighborhood of $O$. Hence 
the integral of the source term of  Eq.~\eqref{eq:forZ} vanishes. 
Since $\del^x Z$ and $\Psi^6 \del^x K$ 
are odd functions of $x$,  we have 
\begin{equation}
\int_{-L}^{+L} \dd x \left(-\frac{4}{3}\del^xZ+\frac{1}{3}\Psi^6\del^xK\right)=0. 
\end{equation}
Hence, the integral of the $x$-component of the source term of Eq.~\eqref{eq:forXi} vanishes. 
The same is true for the other components of Eq.~\eqref{eq:forXi}.

\subsection{Numerical procedure}
\label{sec:numeste}

As shown in the preceding section, 
we have to solve the following three coupled Poisson equations,
\begin{eqnarray*}
\triangle\psi&=&\triangle \left(\frac{M}{2R}W(R)\right)
-\frac{1}{8}(\tilde L X)_{ij}
(\tilde L X)^{ij}\Psi^{-7}
+\frac{1}{12}K^2\Psi^5, \\
\triangle Z&=&\frac{1}{2}\del_i(\Psi^6\del^iK), \\
\triangle X^i&=&-\frac{1}{3}\del^iZ+\frac{2}{3}\Psi^6\del^iK. 
\end{eqnarray*}
In order to get numerical solutions of the above equations, 
we adopt the method of finite differentiations.
By replacing all derivative terms by finite differences, we have a very large 
simultaneous equation. We solve this simultaneous equation by the 
Successive Over-Relaxation method.
 We denote the
values of $\psi$, $Z$ and $X^i$ at each iteration step 
by $\psi_0$, $\psi_1$, $\psi_2$ ..., and so on, 
where the subscript 0 denotes a trial value. 
 At the $(n+1)$-th step of the iteration, the terms corresponding to 
the source terms of the Poisson equations 
are estimated by using $\psi_n$, $Z_n$ and $X^i_n$. 

If we complete the $n$-th step of the iteration, 
we obtain $Z_n$ which satisfies the boundary conditions (\ref{B-1})-(\ref{B-3}). 
Here, we should note that this $Z_n$ does not necessarily 
satisfy the consistency condition \eqref{consistency}. 
In order to obtain $Z_n$ which satisfies Eq.~\eqref{consistency}, 
we can use the degree of freedom to add a constant to $Z$ as follows, 
\begin{equation}
Z\rightarrow Z':=
Z-\frac{1}{L^3}\int_{{\cal D}-\{O\}} \dd x^3 Z. 
\end{equation}
$Z'$ is also a solution of Eq.~\eqref{eq:forZ} and further satisfies 
Eq.~\eqref{consistency},  if $Z$ is a solution of Eq.~\eqref{eq:forZ}. Thus, 
before evaluating the source term, 
we reset the value of 
$Z_n$ as follows: 
\begin{equation}
Z_n\rightarrow Z'_n=Z_n-\frac{1}{L^3}\int_{{\cal D}-\{O\}}\dd x^3Z_n.  
\label{eq:Zzeromode}
\end{equation}

It should also be noted that the boundary conditions already 
given are not enough to close the simultaneous equation, since 
these boundary conditions do not determine homogeneous solutions of the 
Poisson equations for $\psi$ and $X^i$, i.e., their zero modes. 
(The zero mode of $Z$ is already fixed by Eq.~\eqref{eq:Zzeromode}.)
For this purpose, we
need to specify the values of $\psi$
and $X^i$
at one of all numerical grids.
We fix the zero modes of $\psi$ and $X^i$
so that $\psi(0)=1$ and 
$X^i(0)=0$, 
or in other words, 
before evaluating the source terms, 
we add constants to $\psi_n$ and $X^i_n$ as
\begin{eqnarray}
\psi_n(\bm x)&\longrightarrow& \psi'_n(\bm x):=\psi_n(\bm x)-\psi_n(0)+1,\\
X^i_n(\bm x)&\longrightarrow& {X'}^i_n(\bm x):=X^i_n(\bm x)-X_n^i(0). 
\end{eqnarray}
Note that $\psi(0)=1$ is equivalent to the choice of 
$\Psi_{\rm c}=1$ in Eq.~\eqref{Psi-def}. 
Eventually, we evaluate the source terms by using 
$\psi_n'$, $Z'_n$ and ${X'}^i_n$ instead of $\psi_n$, $Z_n$ and ${X}^i_n$. 
The value of $H_{\rm eff}$ is also evaluated through Eq.~\eqref{eq:effHub2} 
by using $\psi_n'$, $Z'_n$ and ${X'}^i_n$ so that 
the integrability condition is satisfied.

\subsection{Results}
\begin{figure}[htbp]
\begin{center}
\includegraphics[scale=1]{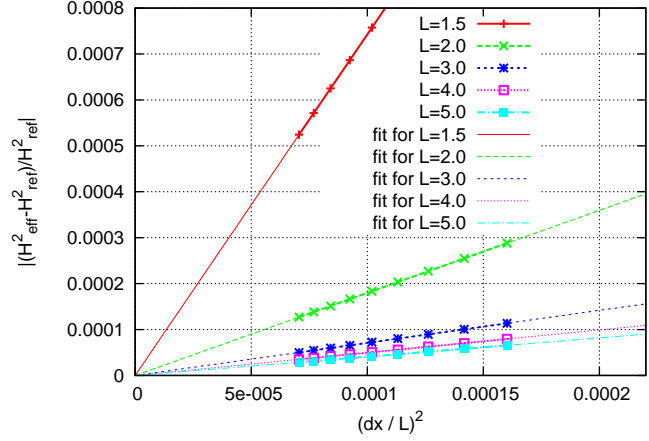}
\caption{Results of convergence test. 
}
\label{fig:conv}
\end{center}
\end{figure}

\begin{figure}[htbp]
\includegraphics[scale=0.8]{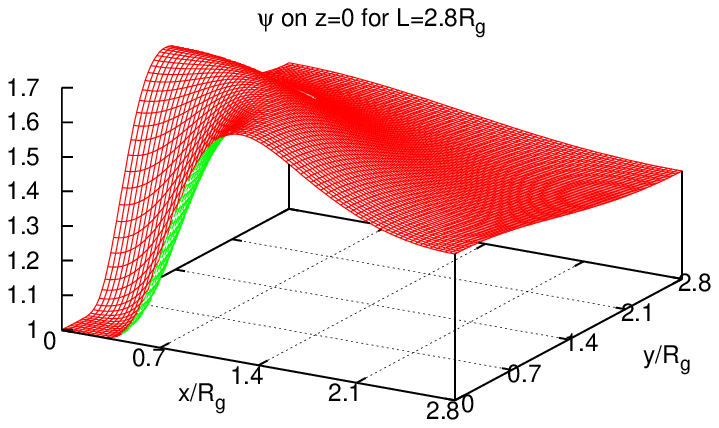}
\hspace{5mm}
\includegraphics[scale=0.8]{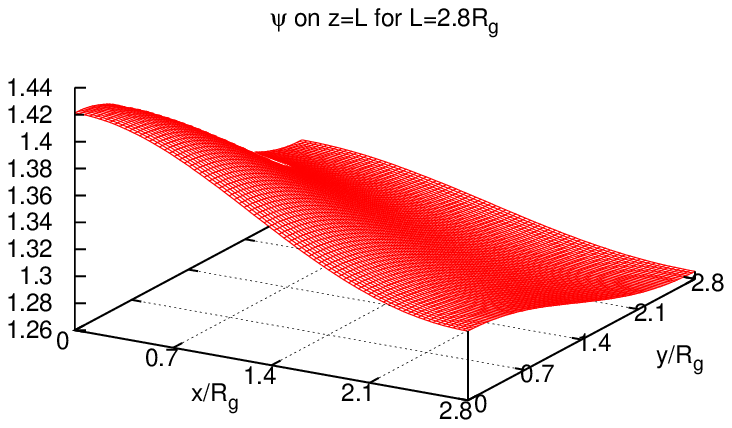}
\vspace{5mm}

\includegraphics[scale=0.8]{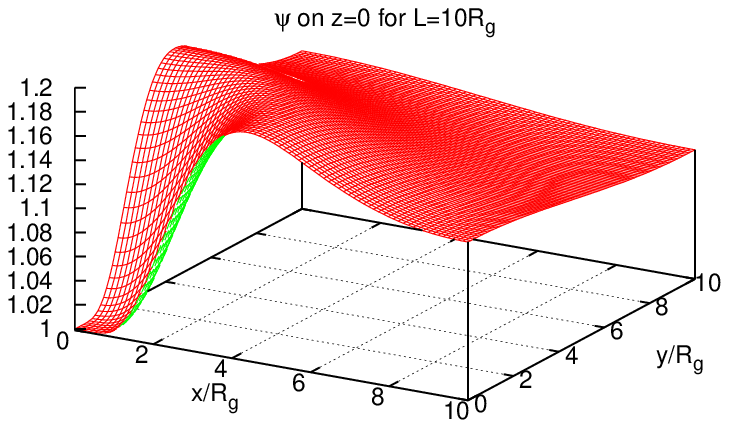}
\hspace{5mm}
\includegraphics[scale=0.8]{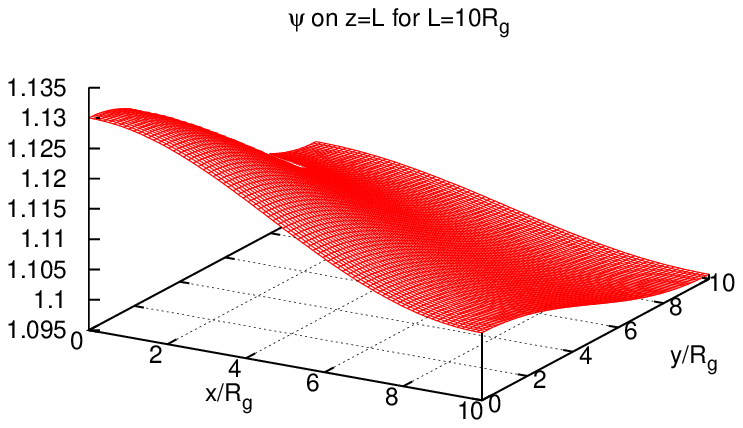}
\caption{
$\psi$ on $z=0$ and $z=L$ 
planes as functions of $x$ and $y$ for $L=2.8R_{\rm g}$ and $L=10R_{\rm g}$. 
}
\label{fig:psi}
\end{figure}

\begin{figure}[htbp]
\includegraphics[scale=0.8]{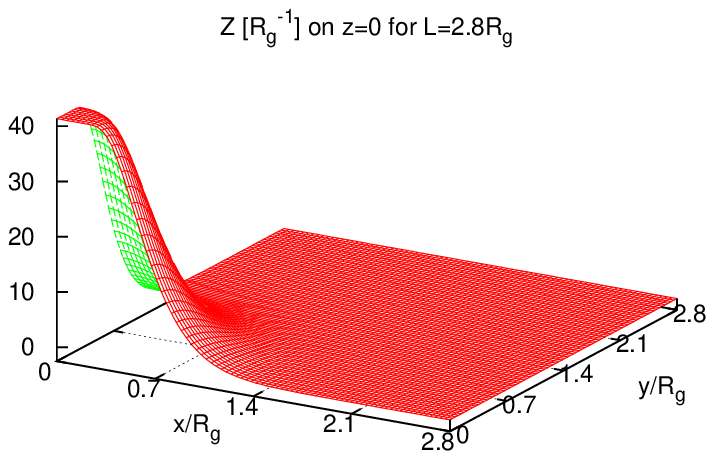}
\hspace{5mm}
\includegraphics[scale=0.8]{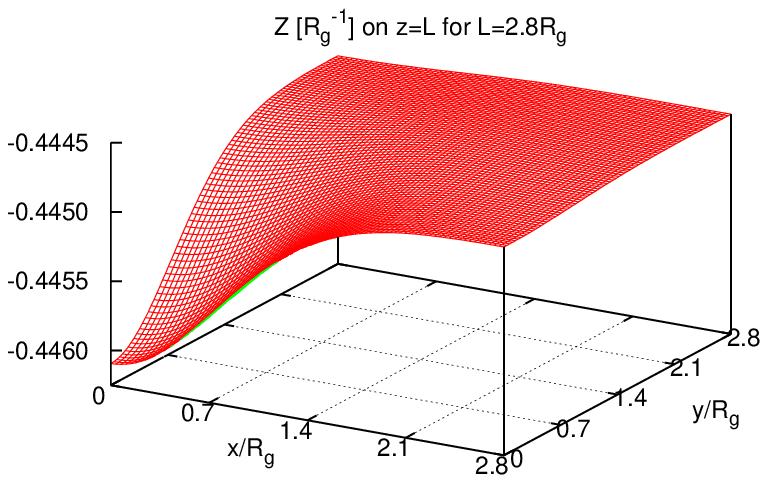}
\vspace{5mm}

\includegraphics[scale=0.8]{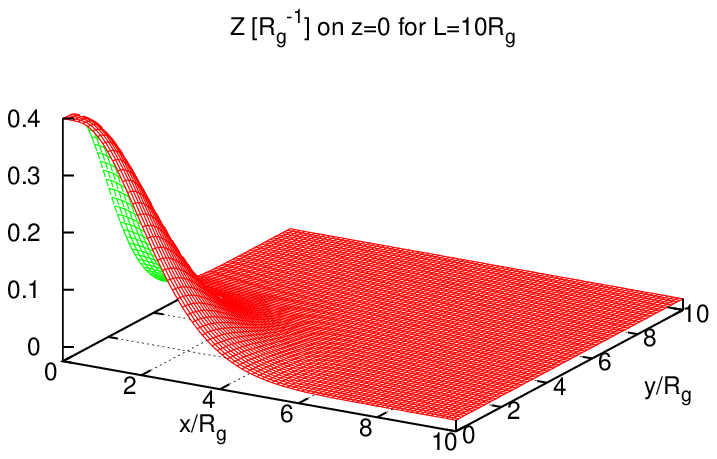}
\hspace{5mm}
\includegraphics[scale=0.8]{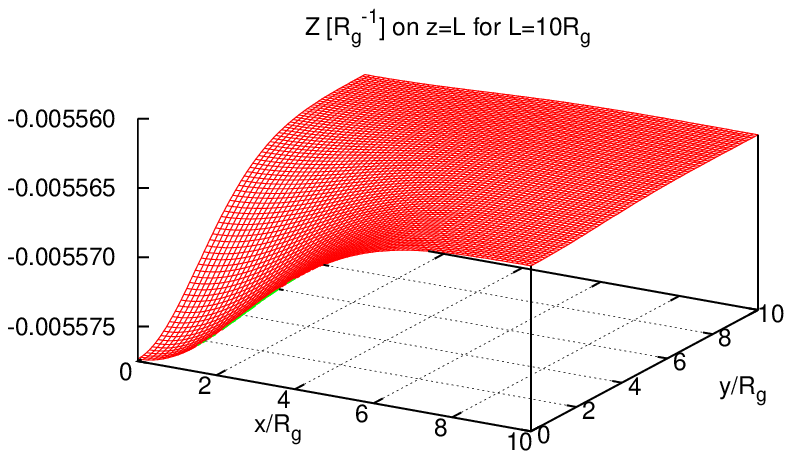}
\caption{
$Z$ on $z=0$ and $z=L$ 
planes as functions of $x$ and $y$ for $L=2.8R_{\rm g}$ and $L=10R_{\rm g}$. 
}
\label{fig:z}
\end{figure}

\begin{figure}[htbp]
\includegraphics[scale=0.8]{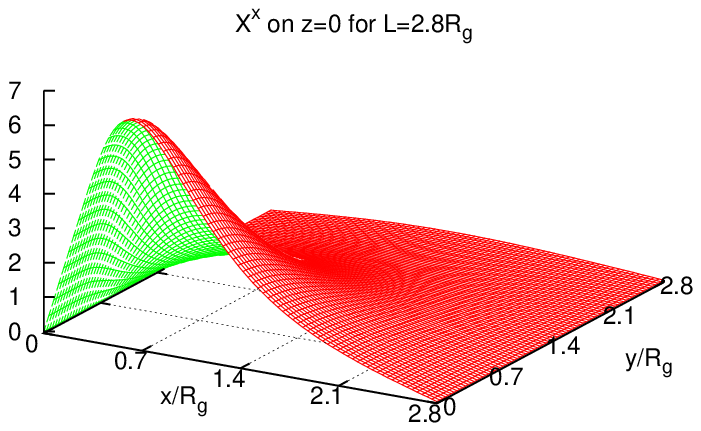}
\hspace{5mm}
\includegraphics[scale=0.8]{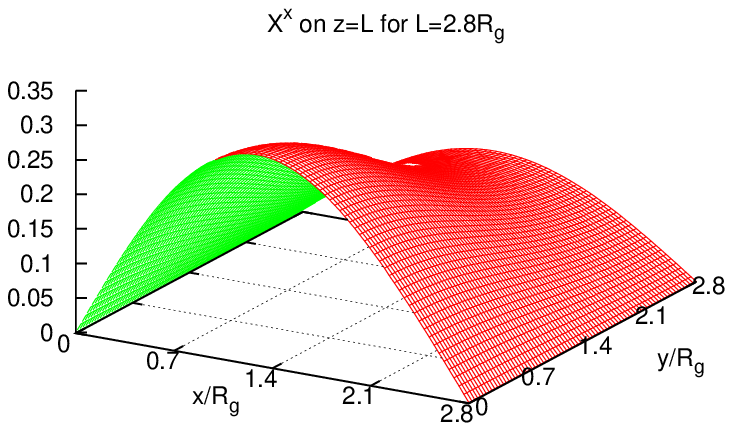}
\vspace{5mm}

\includegraphics[scale=0.8]{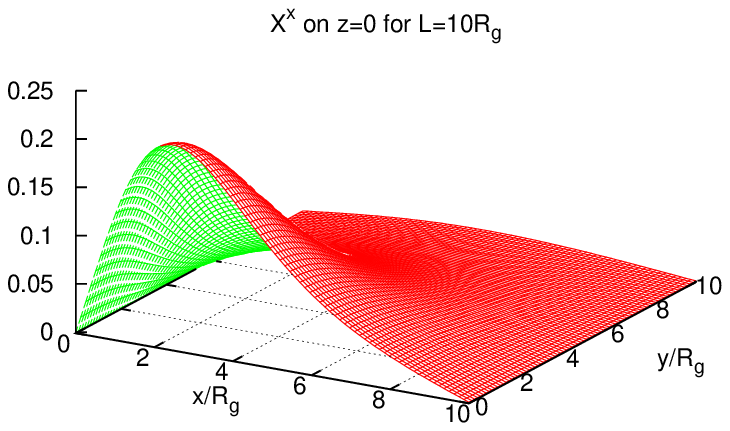}
\hspace{5mm}
\includegraphics[scale=0.8]{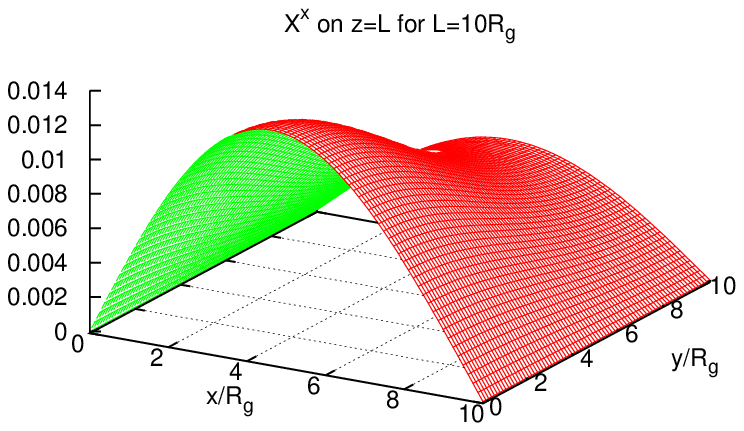}
\caption{
$X^x$ on $z=0$ and $z=L$ 
planes as functions of $x$ and $y$ for $L=2.8R_{\rm g}$ and $L=10R_{\rm g}$. 
}
\label{fig:Xx}
\end{figure}

We solved the constraint equations in the parameter domain  
$2.8R_{\rm g}\leq L\leq 20R_{\rm g}$, where 
\begin{equation}
R_{\rm g}:=\frac{M}{2}.
\end{equation}
As will be shown later, the horizons of a black hole 
are located at $R\simeq R_{\rm g}$. 
The parameters $\sigma$ and $\ell$ which determine  $K$ 
are set to be $\sigma=0.2R_{\rm g}$ and $\ell=L-0.4R_{\rm g}$. 
We could not get convergence for $L$ smaller than $2.8R_{\rm g}$. 
This result implies that there is no solution for $L<2.8R_{\rm g}$ on our assumptions:  
conformally flat metric and no transverse-traceless part of the extrinsic 
curvature.
The results of convergence test 
for each value of $L/R_{\rm g}$ is shown in Fig.~\ref{fig:conv}. 
The second order convergence is confirmed in all cases for 
the value of 
$H_{\rm eff}^2$ 
where the reference value 
$H_{\rm ref}^2$ 
is given by the least-square fit. 
We plot $\psi$, $Z$ and $X^x$ on $z=0$ and $z=L$ 
planes
as functions of $x$ and $y$ for $L=2.8R_{\rm g}$ and $L=10R_{\rm g}$ in 
Figs.~\ref{fig:psi}, \ref{fig:z} and \ref{fig:Xx}. 

\section{Analysis of the initial data}
\label{sec:analysis}
\subsection{Inhomogeneities}
\label{sec:inhomogeneity}
\begin{figure}[htbp]
\begin{center}
\includegraphics[scale=0.8]{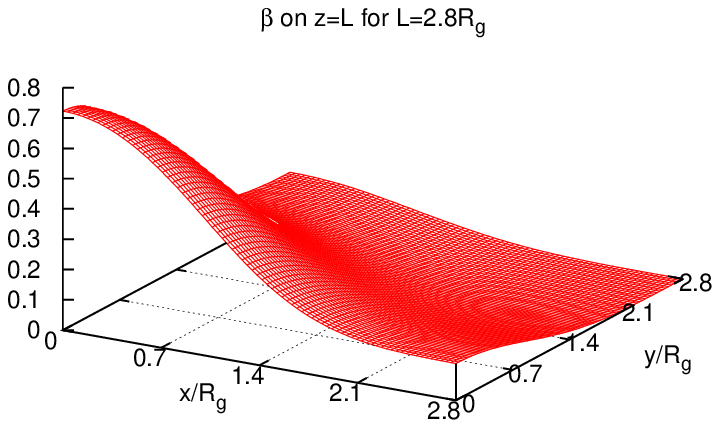}
\hspace{5mm}
\includegraphics[scale=0.8]{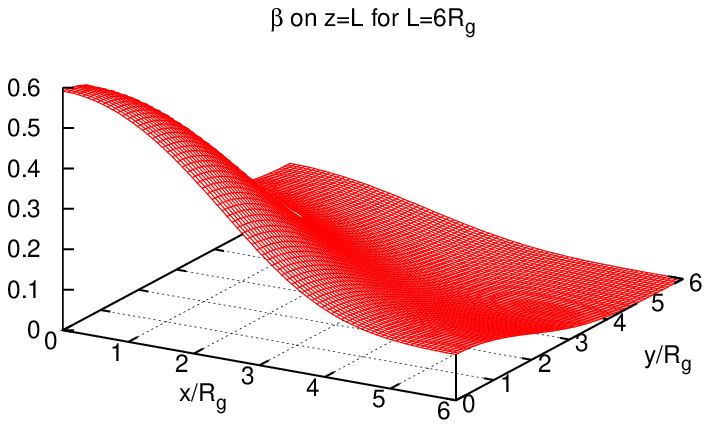}
\vspace{5mm}

\includegraphics[scale=0.8]{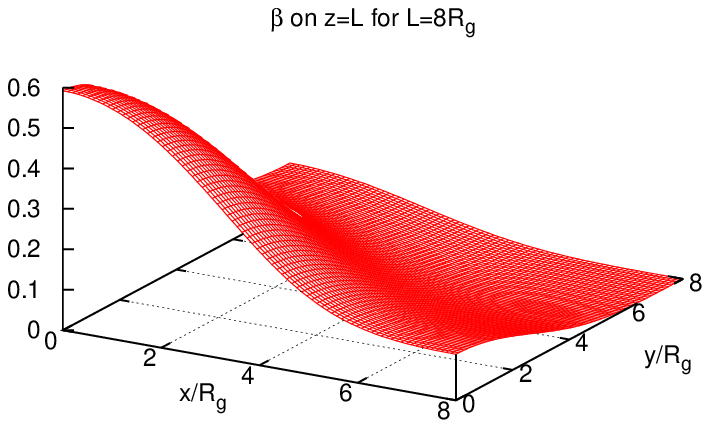}
\hspace{5mm}
\includegraphics[scale=0.8]{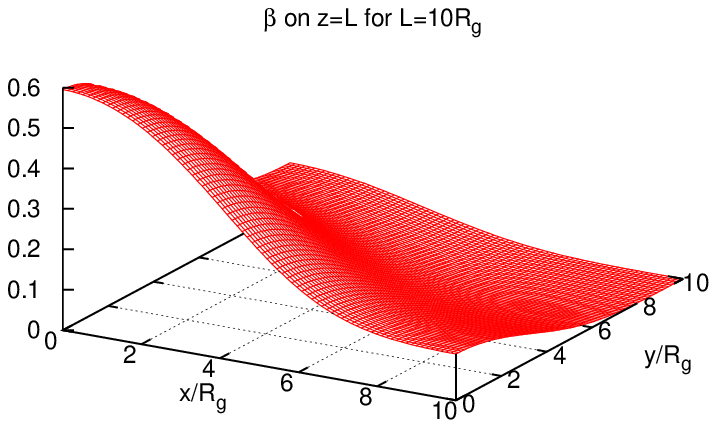}
\caption{
$\beta$ on $z=0$ and $z=L$ 
planes as functions of $x$ and $y$ for $L=2.8R_{\rm g}$ and $L=10R_{\rm g}$. 
}
\label{fig:trlricci}
\end{center}
\end{figure}

First, we demonstrate the inhomogeneities of our initial data. 
For this purpose, we 
investigate
the following quantity:
\begin{equation}
\beta:=\frac{\gamma^{ac}\gamma^{bd}\mathcal R^{\rm T}_{ab}\mathcal R^{\rm T}_{cd}}
{\gamma^{ik}\gamma^{jl}\mathcal R_{ij}\mathcal R_{kl}}, 
\end{equation}
where $\mathcal R_{ij}$ and $\mathcal R^{\rm T}_{ij}$ denote the 3-dimensional Ricci curvature tensor and 
its traceless part, respectively. 
We use $\beta$ as a measure of homogeneity 
and isotropy,  
since 
a region with $\beta=0$ 
and 
$\partial_i \mathcal R=0$ 
is homogeneous and isotropic.
Since we are interested in the inhomogeneities far from black holes, we plot the value 
of $\beta$ on $z=L$ plane, which is one of the faces of the domain $\mathcal D$, 
as a function of $x$ and $y$ in Fig.~\ref{fig:trlricci}. 
The quantity $\beta$ almost vanishes in the vicinity of a vertex $x=y=z=L$. 
Further, the norm of the traceless part of the extrinsic curvature 
$\Psi^{-12}(\tilde{L}X)_{ij}(\tilde{L}X)^{ij}$ is much less than the 
square of the trace part of the extrinsic curvature $K^2$ in the 
neighborhoods of the vertices.  We find from the Hamiltonian constraint 
together with this fact that $\mathcal R \simeq -K^2=$constant 
and hence $\mathcal R_{ij}\simeq -3H_{\rm eff}^2\gamma_{ij}$,  
in these regions. Thus, the neighborhoods of the vertices are 
well approximated by the Milne universe model which is the 
Minkowski spacetime foliated by the family of homogeneous 
and isotropic spacelike hypersurfaces 
with negative  Ricci curvature scalar.
Conversely, around the center of a 
face of ${\cal D}$ 
($x=y=0$ and $z=L$), 
the inhomogeneity remains even 
if $L \gg R_{\rm g}$. 
We may understand this result as follows. If the neighborhoods of all faces of 
$\mathcal D$ were well approximated by the Milne universe 
model, a 3-hyperboloid would be tiled with 
the lattice structure shown in Fig.~\ref{fig:cube}.
However, this consequence conflicts with a 
mathematical theorem about ``tiling"\cite{Clifton:2009jw,Coxeter}. 
Therefore, our initial data cannot be 
homogeneous and isotropic in the neighborhoods of all faces 
of $\mathcal D$ even for $L \gg R_{\rm g}$ 
as is explicitly shown in Fig.~\ref{fig:trlricci}. 

\subsection{Horizons}
\label{sec:horizons}
We define a horizon as a spacelike closed 2-surface  
with vanishing expansion of a null vector field normal to the 2-surface. 
There are two independent null directions normal 
to the 2-surface, so there are two kinds of horizons accordingly.  
Here, we consider these horizons in 
the domain $\mathcal D-\{O\}$. 
A closed 2-surface divides the domain $\mathcal D-\{O\}$ into two regions. 
In this paper, since we are interested in the horizons associated with a black hole, 
we focus on a case in which one of the two regions includes 
the puncture.  
We call the domain including the puncture the {\it inside}, 
whereas the other domain is called the {\it outside}. 
Then, we call the direction from a point on a 
closed 2-surface
to 
the outside the {\it outgoing} direction, 
whereas the opposite direction is called 
{\it ingoing} direction. 
Accordingly, we name a horizon with 
vanishing expansion of the outgoing null vector field 
the black hole (BH) horizon, whereas a horizon with vanishing expansion of the 
ingoing null vector field is named the white hole (WH) horizon. 

The expansions of the null vector fields normal 
to this 2-surface are given by 
\begin{equation}
\chi_\pm=(\gamma^{ij}-s^is^j)(\pm D_is_j-K_{ij}), 
\end{equation}
where the subscript $+$ means that of the outgoing null, whereas 
the subscript $-$ represents that of the ingoing null, and 
$s^i$ is the outgoing unit vector which is normal 
to this 2-surface 
and
tangent to the initial hypersurface.  
Defining $\tilde s^i$ and $\tilde s_i$ as 
\begin{equation}
\tilde s^i:=\psi^2 s^i~,~~\tilde s_i:=\delta_{ij}\tilde s^j, 
\end{equation}
we rewrite $\chi_{\pm}$ in the form
\begin{equation}
\chi_\pm=(\tilde s_i\tilde s_j-\delta_{ij})
\left[\Psi^{-6}(\tilde LX)^{ij}+\frac{1}{3}\delta^{ij}K\right]
\pm \Psi^{-2}\del_i \tilde s^i
\pm4\Psi^{-2}\tilde s^i
\del_i\ln\Psi. 
\end{equation}

\begin{figure}[htbp]
\begin{center}
\includegraphics[scale=0.8]{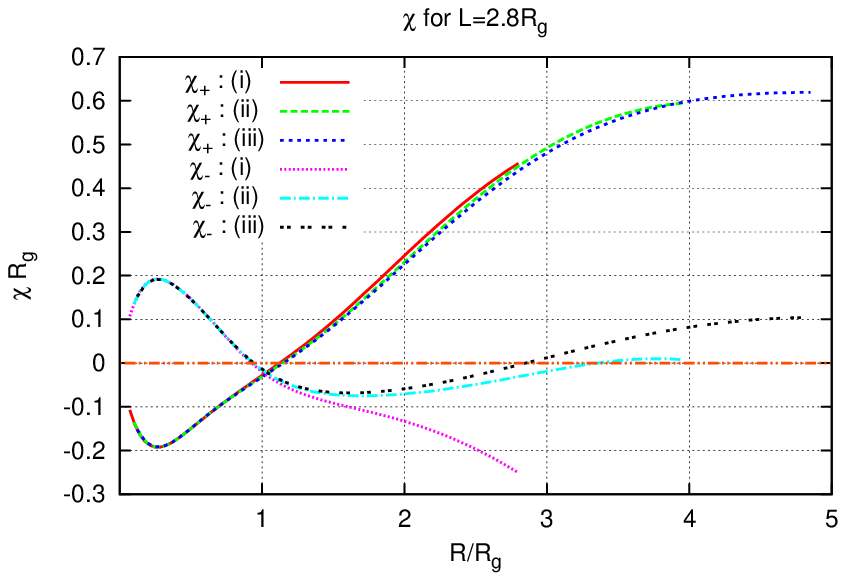}
\hspace{5mm}
\includegraphics[scale=0.8]{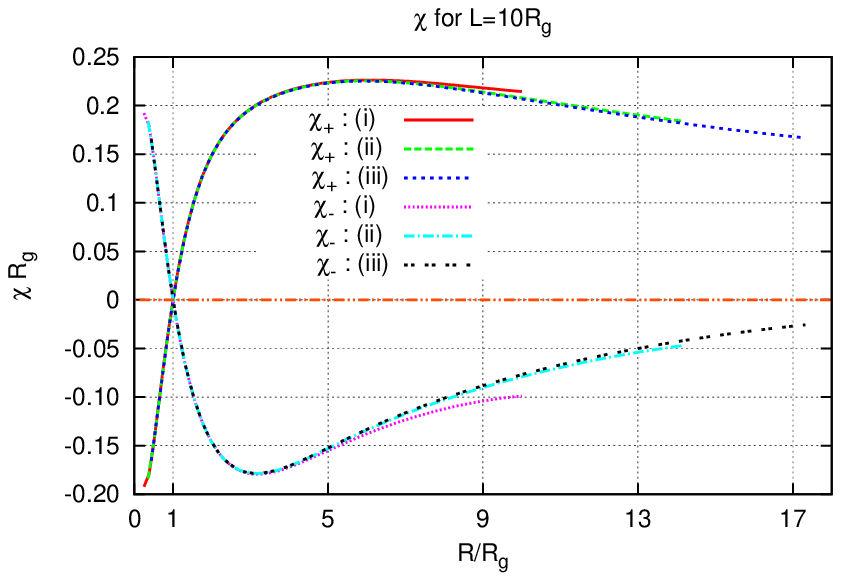}
\caption{$\chi_\pm$ as functions of $R$ for $L=2.8R_{\rm g}$ and $L=10R_{\rm g}$. 
}
\label{fig:expansion}
\end{center}
\end{figure}

In this paper, 
in stead of solving the equation $\chi_\pm=0$, 
we investigate the expansions of the null vector fields normal to various spheres 
centered at the origin $O$. 
The conformal unit vector $\tilde{s}^i$ normal to the sphere of the radius $R$ is given by%
\begin{equation}
\tilde s^i= \frac{x^i}{R}. 
\label{eq:sidef}
\end{equation}
If the initial hypersurface is almost spherically symmetric near the horizon, 
the horizon is also almost spherically symmetric and 
$\tilde s^i$ 
is a good approximation of the unit vector field normal to the horizon.
In Fig.~\ref{fig:expansion}, we plot the expansions $\chi_\pm$ as functions of $R$ on 
the following three lines:
\begin{eqnarray*}
&{\rm (i)}&y=0,~z=0, \\
&{\rm (ii)}&x=y,~z=0, \\
&{\rm (iii)}&x=y=z. 
\end{eqnarray*}

From these figures, we see that 
there are spheres which are very good approximations of horizons; 
the expansion $\chi_+$ or $\chi_-$ at the 
intersections of these spheres and the lines 
(i) - (iii) vanishes. 
The coordinate 
radius $R$ of the BH horizon is 
 equal to $1.14R_{\rm g}$ 
in the case of $L=2.8R_{\rm g}$, whereas it is 
equal to $R_{\rm g}$ 
in the case of $L=10R_{\rm g}$. 
The coordinate radius $R$ of the 
WH horizon is 
equal to $0.92R_{\rm g}$
in the case of $L=2.8R_{\rm g}$, whereas it is 
equal to $R_{\rm g}$ in the case 
of $L=10R_{\rm g}$. 
In the case of $L=2.8R_{\rm g}$, 
the WH horizon is located inside the BH horizon. 
Since the domain $R\lesssim R_{\rm g}$ 
is well approximated by the Schwarzschild BH, 
we can say  
that the initial hypersurface is passing through the future
of the bifurcation point of the horizons for $L=2.8R_{\rm g}$. 
On the other hand, in the larger $L$ cases, 
since
the WH and BH horizons coincide with each other, 
we may say that the initial hypersurface
is passing through a domain very close to the bifurcation point. 

\subsection{Effective Hubble %
equation}
\label{sec:averaging}

The mass density of black holes $\rho_{\rm H}$  defined by Eq.~\eqref{eq:rhoBH} 
is roughly estimated at about $M/8L^3$. 
If the kinetic energy density  $\rho_{\rm K}$ 
defined by \eqref{eq:rhoK} is much less than 
$\rho_{\rm BH}$, the effective Hubble parameter 
$H_{\rm eff}$ is roughly estimated at about 
$H_{\rm eff}^2\sim 8\pi\rho_{\rm BH}/3\sim \pi M/3L^3$. 
Then, 
in the covering space of the domain $\mathcal D-\{O\}$, 
the number $N_{\rm BH}$ of black holes within a 
sphere of the 
cosmological
horizon radius 
$H_{\rm eff}^{-1}$ is about
\begin{equation}
N_{\rm BH}\sim \frac{1}{M} \times\frac{4\pi}{3}H_{\rm eff}^{-3}\rho_{\rm BH}
\sim\frac{1}{4}\left(\frac{3L^3}{2\pi R_{\rm g}^3}\right)^{1/2}.
\end{equation}
If $L/R_{\rm g}$ is much larger than unity, there are many black holes within a sphere of the cosmological
horizon radius, and thus the black hole universe 
would be very similar to the Einstein-de Sitter(EdS) universe.

\begin{figure}[htbp]
\begin{center}
\includegraphics[scale=1.5]{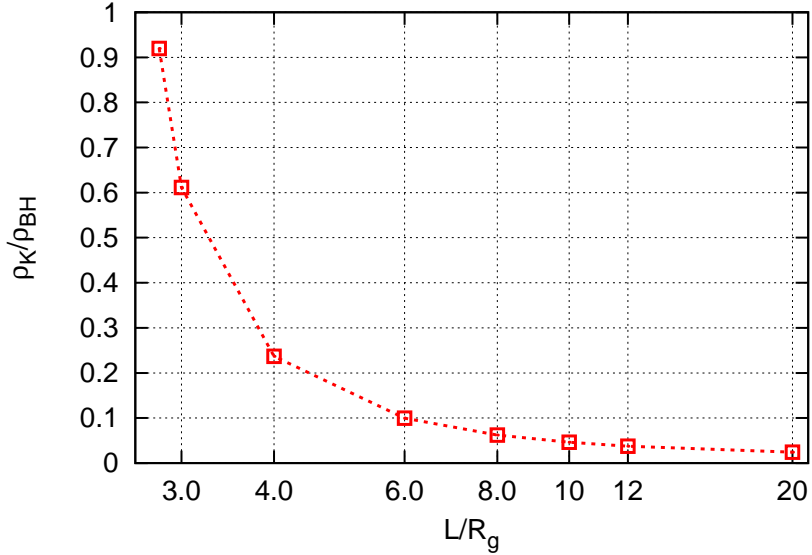}
\caption{
$\rho_{\rm K}/\rho_{\rm BH}$ 
as a function of $L/R_{\rm g}$. 
}
\label{fig:LXpart}
\end{center}
\end{figure}

From the above consideration, we expect that 
the effective Hubble parameter and the 
mass density 
of black holes
asymptotically satisfy the Hubble equation 
of the EdS universe 
in the limit of $L/R_{\rm g}\rightarrow\infty$. 
That is, we expect that 
the effective Hubble parameter 
 behaves asymptotically as 
\begin{equation}
H_{\rm eff}^2 \longrightarrow \frac{8\pi}{3}\rho_{\rm BH}. 
\label{eq:asymH}
\end{equation}
This means that the contribution of $\rho_{\rm K}$ 
decreases with larger $L/R_{\rm g}$. 
We depict $\rho_{\rm K}/\rho_{\rm BH}$
as a function of $L/R_{\rm g}$ in Fig.~\ref{fig:LXpart}.  
It is seen from this figure that 
$\rho_{\rm K}/\rho_{\rm BH}$
asymptotically vanishes for large $L/R_{\rm g}$ 
and the effective Hubble equation approaches that of EdS universe. 

It is suggestive to regard the one parameter family of 
the initial data sets as a fictitious time evolution 
of the black hole universe. 
Eq.~\eqref{eq:effHub2} gives the effective Hubble parameter
at each time of the fictitious evolution. 
If we define an effective scale factor by 
\begin{equation}
a_V:=V^{1/3}, 
\end{equation}
Eq.~\eqref{eq:asymH} means that 
$H_{\rm eff}^2$ asymptotically behaves as $\propto 1/a_V^3$ 
when the universe expands enough. 

\begin{figure}[htbp]
\begin{center}
\includegraphics[scale=1]{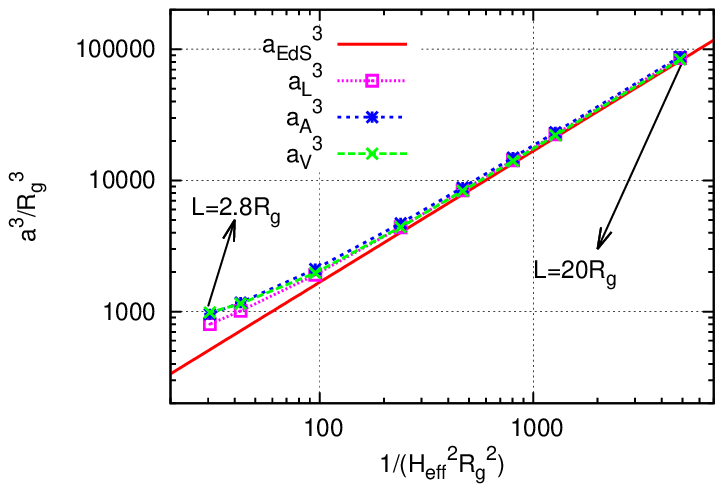}
\hspace{5mm}
\includegraphics[scale=1]{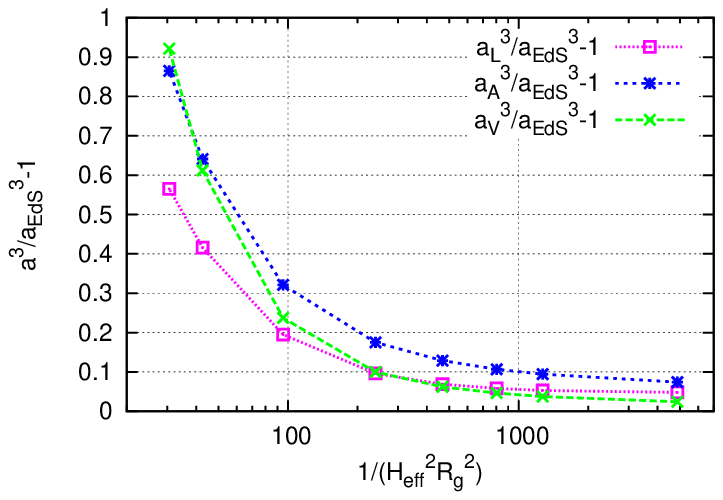}
\caption{
Effective scale factors(left panel) and deviations of them 
from the fiducial scale factor $a_{\rm EdS}$(right panel) as 
functions of the effective Hubble parameter. 
}
\label{fig:scalefac}
\end{center}
\end{figure}

Other remarkable ways to define effective scale factors 
are to use the proper area of the boundary and the proper length 
of the edge of the cubic domain $\mathcal D$. 
Let $a_A$ and $a_L$ denote the effective scale factors 
defined by using the proper area and the edge length, respectively. 
$a_L$ is defined by the proper length of a edge itself and 
$a_A$ is defined by 
\begin{equation}
a_A:=\sqrt{\frac{A}{6}}, 
\end{equation}
where $A$ is the proper area of $\del \mathcal D$. 
In addition, we define the fiducial scale factor $a_{\rm EdS}$
by using the Hubble equation of EdS universe as follows:
\begin{equation}
a^3_{\rm EdS}:=\frac{8\pi}{3H_{\rm eff}^2}. 
\end{equation}
The relation between effective scale factors and the effective Hubble 
parameter is shown in Fig.~\ref{fig:scalefac}. 
All effective scale factors asymptotically behave as 
$\propto H_{\rm eff}^{-2/3}$ for larger $L/R_{\rm g}$, that is, 
the behaviour of the effective Hubble parameter 
as a function of an effective scale factor agrees with 
that of the EdS universe 
at late time of the fictitious time evolution. 
We note that, even though all effective scale factors 
are asymptotically proportional to $H_{\rm eff}^{-2/3}$, 
the proportionality coefficients are different from each other. 
It seems that the proportionality coefficient for $a_V$ 
asymptotically agrees with that for $a_{\rm EdS}$, but 
it is not true for $a_A$ and 
$a_L$(see the right panel of Fig.~\ref{fig:scalefac}).

\subsection{Comparison with the Newtonian approximation and backreaction effect}
\label{sec:newton}
One possible way of approximation which considerably reduces 
the numerical effort is the cosmological Newtonian approximation. 
The cosmological $N$-body simulation based on this approximation scheme 
is very useful to study the structure formation in the universe indeed. 
The $N$-body simulation follows the motion of point particles 
gravitationally interacting with each other, and these particles are regarded as 
the dark matter in the cosmological context.  
The black hole is a candidate for the ingredient of the dark matter in our universe. 
However, since the black hole is a highly relativistic object, it is non-trivial 
whether the dark matter composed of black holes is well described 
by the cosmological $N$-body simulation based on the Newtonian approximation. 
Our black hole universe 
model is a relativistic version of the cosmological $N$-body system, and  
thus, by using this model, we can see in what situation the Newtonian 
$N$-body simulation correctly describes the motion of the dark matter composed of 
black holes. 

In the cosmological Newtonian approximation scheme, 
the gravitational force 
is given by the spatial gradient of the 
Newtonian potential $\Phi$ which is related to 
the conformal factor $\Psi$ by
$1-2\Phi =\Psi^4$, and thus 
the Newtonian potential is obtained by solving the Hamiltonian constraint, 
which gives the Hubble equation after averaging. 
Since the metric is assumed to be almost 
equal to that of the EdS universe model, 
the term proportional to $(\tilde{L}X)_{ij}(\tilde{L}X)^{ij}$ 
should be so small that it is a negligible higher order correction 
in the Hamiltonian constraint. 
Hence, we do not need to solve the momentum constraint. 

Before considering a point particle as 
the ingredient of the $N$-body simulation, 
we assume that the particle is a spherical ball with the 
finite energy density $\rho(\bm{x})$. 
Further, we assume the similar situation to our black hole universe; 
the particle 
has the mass $M$, the center of the particle is 
located at the origin $O$ in the cubic domain $\cal D$ 
whose edge length is $2L$, and the periodic boundary condition is imposed. 
By definition, we have
\begin{equation}
M=\int_{\cal D}\rho(\bm{x})d^3\bm{x}.
\end{equation}
The time slicing condition up to the Newtonian order is assumed to be \begin{equation}
K=-3H_{\rm N},
\end{equation}
where $H_{\rm N}$ is the effective Hubble parameter up to the 
Newtonian order and is determined by 
\begin{equation}
H_{\rm N}^2=\frac{8\pi}{3}\times \frac{M}{8L^3}. \label{N-Hubble}
\end{equation}
Here note that $H_{\rm N}$ is the same as 
the Hubble parameter of the background EdS 
universe model.  
Then, since nonlinear terms with respect to $\Psi$ 
in the Hamiltonian constraint is linearized with 
respect to $\Phi$, the Hamiltonian constraint 
takes the following form in the cosmological Newtonian 
scheme\cite{1980lssu.book.....P,Shibata:1995dg}; 
\begin{equation}
\triangle \Phi=4\pi \left[\rho(\bm{x})-\frac{M}{8L^3}\right].  \label{Newton-1}
\end{equation}
In the cosmological Newtonian approximation scheme, 
$\rho$ can be much larger than $M/8L^3$, 
but $\rho$ should be so small that $|\Phi|$ is much smaller than unity.   

Let us consider the case in which 
the size of the particle is much smaller than $L$. 
In this case, 
since the tidal force can be neglected, 
it is enough to 
consider the energy density for a point particle given by 
$M\delta(\bm{x})$ instead of the finite energy density 
$\rho(\bm{x})$. 
Using this approximation, 
we can accurately 
estimate the gravitational 
force produced by a particle at points of other particles. 
%
Then the  Hamiltonian constraint in the cosmological $N$-body system is given by
\begin{equation}
\triangle \Phi=4\pi M \delta(\bm x)-\frac{\pi M}{2L^3}.  \label{Newton-2}
\end{equation}
Equation \eqref{Newton-2} is the basic equation for 
the cosmological $N$-body simulation 
based on the Newtonian approximation scheme. 

In our case, 
since the $\Psi$ diverges at $O$ 
in the black hole universe, 
it is obvious that 
the cosmological Newtonian approximation is not 
applicable in whole region of $\mathcal D-\{O\}$. 
%
%
When we compare the black hole universe with 
the cosmological Newtonian system given by 
Eq.~\eqref{Newton-2}, 
%
the point-particle approximation, i.e., $\rho(\bm{x})=M\delta(\bm{x})$,  
should be regarded as a technical simplification. 
Hence, it is a very non-trivial issue whether  
a point particle in the cosmological Newtonian $N$-body system may be 
identified with a black hole.

In order to numerically obtain solutions 
of Eq.~\eqref{Newton-2}, we decompose $\Phi$ as follows,
\begin{equation}
\Phi=\phi-\frac{M}{R}\left[1-W(R)\right]. 
\end{equation}
The equation for $\phi$ is given by
\begin{equation}
\triangle \phi=-\triangle\left(\frac{M}{R}W(R)\right)-\frac{\pi M}{2L^3}. 
\end{equation}
This can be numerically integrated in $\mathcal D-\{O\}$ by using the same 
way as in Sec.~\ref{sec:numeste}. 

\begin{figure}[htbp]
\begin{center}
\includegraphics[scale=0.8]{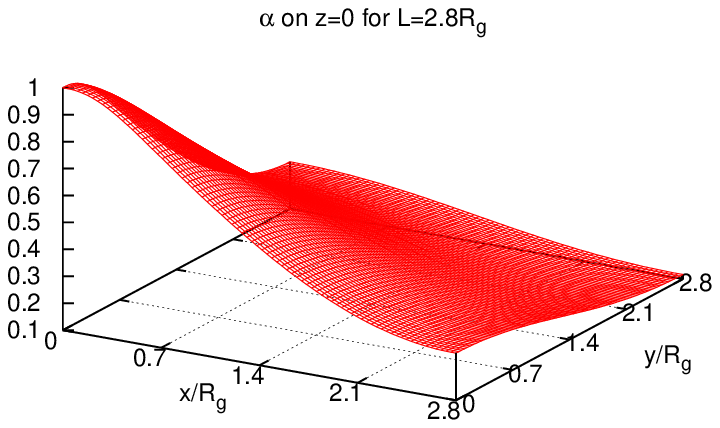}
\hspace{5mm}
\includegraphics[scale=0.8]{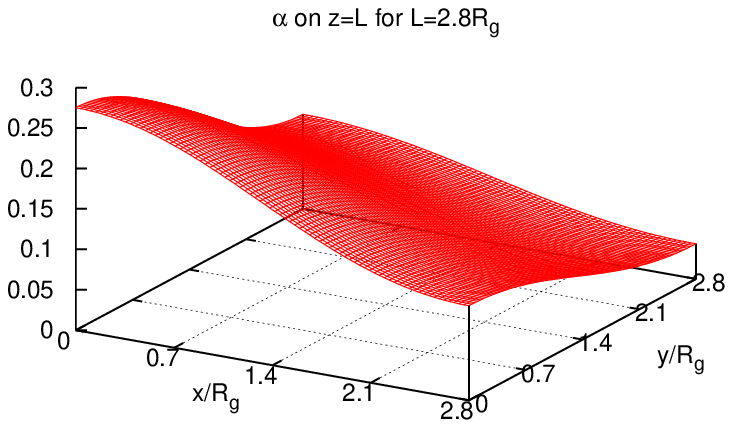}
\vspace{5mm}

\includegraphics[scale=0.8]{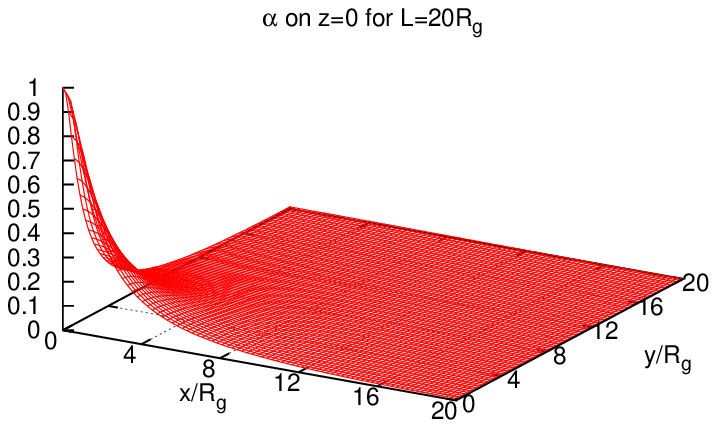}
\hspace{5mm}
\includegraphics[scale=0.8]{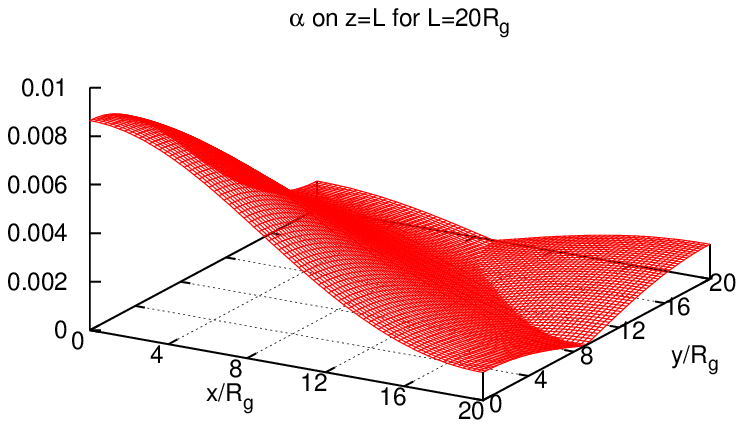}
\caption{
$\alpha$ on $z=0$ and $z=L$ surfaces for $L=2.8R_{\rm g}$ and $L=20R_{\rm g}$. 
}
\label{fig:res}
\end{center}
\end{figure}

To show the deviation of the solution obtained by the 
cosmological Newtonian approximation 
from the corresponding relativistic one, we plot the following quantity: 
\begin{equation}
\alpha:=\left|\frac{\Psi^4-1+2\Phi}{\Psi^4}\right|. 
\end{equation}
In Fig.~\ref{fig:res}, $\alpha$ is plotted 
as a function of the coordinates $x$ and $y$ on $z=0$ and $z=L$ 
planes for $L=2.8R_{\rm g}$ and $L=20R_{\rm g}$, respectively. 
We can see that while the deviation around the boundary of 
$\mathcal D$ is a few tens of percent 
for $L=2.8R_{\rm g}$, it is less than 1\% 
for $L=20R_{\rm g}$. 
This result implies that the cosmological Newtonian approximation predicts 
the spatial metric around the boundary of $\mathcal D$ 
very accurately for $L\geq20 R_{\rm g}$. 

%
As mentioned, the effective Hubble parameter $H_{\rm N}$ 
defined by Eq.~\eqref{N-Hubble}
agrees with that of the background Einstein-de Sitter 
universe model. The so-called {\em backreaction} effect is the change of the 
Hubble parameter from the background value 
due to the nonlinear effect of the inhomogeneities. 
Thus, in the present case, we call the effect which causes difference 
between the full relativistic Hubble parameter $H_{\rm eff}$ and 
the background value $H_{\rm N}$  the backreaction effect. 
%
%
%

\begin{figure}[htbp]
\begin{center}
\includegraphics[scale=1]{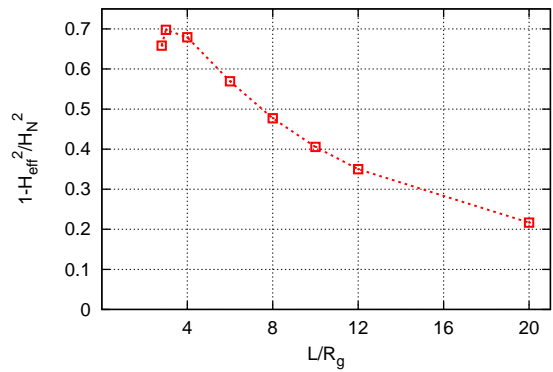}
\caption{$1-H_{\rm eff}^2/H_{\rm N}^2$ as a function of $L/R_{\rm g}$. 
}
\label{fig:hubble}
\end{center}
\end{figure}

To see the significance of the backreaction effect, we 
compare  $H_{\rm eff}^2$ to $H_{\rm N}^2$ 
with fixed $L/R_{\rm g}$. 
As a result of the numerical investigation, we find that 
$H_{\rm N}^2$ has about 20\% deviation 
from $H_{\rm eff}^2$ even for $L=20R_{\rm g}$. 
We plot the value of $1-H_{\rm eff}^2/H_{\rm N}^2$ as a function of 
$L/R_{\rm g}$ in Fig.~\ref{fig:hubble}. 
It is worthwhile to notice that the Newtonian Hubble 
parameter is larger than the relativistic one. 
This means that the backreaction effect 
acts as the brake in the black hole universe model. 
Further, our result means that, 
in the case of $L\leq20R_{\rm g}$, the backreaction effect is 
so large that the cosmological Newtonian 
approximation cannot predict correctly 
the global cosmic volume expansion rate.  
However, Fig.~\ref{fig:hubble} suggests that 
the deviation of $H_{\rm N}$ from 
$H_{\rm eff}$ 
decreases with larger $L/R{\rm g}$,
and hence it seems that the Newtonian $N$-body 
simulation becomes correct asymptotically 
for $L/R_{\rm g}\rightarrow \infty$. 

As already shown, in the case of $L=20R_{\rm g}$, 
the relative difference in the spatial metric 
between the Newtonian scheme and the full relativistic one 
is a few percents on the boundary of $\cal D$, 
and hence the relative differences in the length of an edge and the 
area of a face are also a few percents.  
Furthermore, $\rho_{\rm K}$ defined by Eq.~\eqref{eq:rhoK} 
is about 2\% of  $\rho_{\rm BH}$ (see Fig.~\ref{fig:LXpart}). 
Thus, the difference between $H_{\rm eff}^2$ and $H_{\rm N}^2$ 
comes from the difference between the volume $V$ defined by Eq.~\eqref{V-def} and $8L^3$; 
$V$ is about 1.3 times  larger than $8L^3$. 

Here, we should note that the backreaction effect is large even in the case of $L=20R_{\rm g}$, 
but, as shown in the preceding section, 
the expansion law of the black hole universe model 
might be almost the same as that of the EdS universe. These results would imply that 
the backreaction effect in the black hole universe model would not change the expansion 
law from the EdS universe model but apparently shifts the time to the future. 
However, in order to get definite conclusion, the investigation of the time evolution 
is necessary. 
%

\section{Summary and Conclusion}
\label{sec:summary}
In this paper, we have 
constructed numerically the 
initial data 
of an expanding universe 
model
which is composed of regularly aligned black holes. 
This system is equivalent to a black hole located at the center $O$ of a cubic 
domain $\mathcal D$ with periodic boundary conditions. The black hole is 
represented by a structure like the Einstein-Rosen bridge, and thus 
$O$ corresponds to the asymptotically flat spatial infinity. Since 
the physical domain does not include infinity,  
the physical domain is $\mathcal D$ with 
$O$ removed, i.e., 
$\mathcal D-\{O\}$ whose topology is ${\bf T}^3$ with one point removed. 
The functional form of the trace of the extrinsic curvature $K(\bm x)$
has been chosen 
so that $K$ is a negative constant denoted by $-3H_{\rm eff}$ in the neighborhoods of the faces of 
$\mathcal D$ and vanishes in the neighborhood of $O$, 
where $H_{\rm eff}$ corresponds to 
the effective Hubble parameter.
These requirements are compatible with 
a finite expansion rate of the universe and 
the puncture method 
to numerically treat a black hole,
respectively. Then, we can solve constraint equations 
by giving
the parameter $L/R_{\rm g}$, where $L$ is 
the coordinate length of an edge of the cubic domain $\mathcal D$, 
and $R_{\rm g}$ 
gives a coordinate value which 
is almost equal to the coordinate radius of 
the black hole horizon.
The value of $H_{\rm eff}$ is determined so that the integral of the Hamiltonian 
constraint over $\mathcal D-\{O\}$  is compatible with 
the periodic boundary conditions; this integral leads to the 
effective Hubble equation. 
We find from numerically obtained solutions 
that the neighborhoods of vertices of 
$\mathcal D$ are well approximated by the Milne universe, 
whereas the other region remains inhomogeneous 
even in the case of $L\gg R_{\rm g}$. 
This result implies that the initial data 
of the black hole universe model is inhomogeneous 
even near the faces of $\mathcal D$ 
irrespective of the value of $L/R_{\rm g}$. 

We could  find one white hole and one black hole horizons 
in the present initial hypersurface of $\mathcal D -\{O\}$, 
and both are almost spherically symmetric.
This result implies that the region $R\lesssim R_{\rm g}$ is well approximated by the 
Schwarzschild black hole, and  the initial hypersurfaces considered here are 
passing through 
the future of the bifurcation point of the horizons or 
a very close point to it. 

In order to compare our initial data with 
the Einstein-de Sitter(EdS) universe, 
we studied the relation between the effective mass density $\rho_{\rm BH}$  
of black holes and the effective Hubble parameter $H_{\rm eff}$, which are 
defined in a simple and natural way.
Then, our numerical solutions imply that $\rho_{\rm BH}$ and $H_{\rm eff}$ 
asymptotically satisfy the Hubble equation 
of the EdS universe for $L\gg R_{\rm g}$. 
Once we regard our one parameter family of initial data sets 
as fictitious time evolution of the black hole universe, 
our result would imply that the Hubble equation of the EdS universe 
would be realized when the universe expands enough. 

The validity of the Newtonian approximation in the system 
has also been discussed. 
We numerically solved 
the Hamiltonian constraint equation simplified by 
the cosmological Newtonian approximation and 
compared it with the full solution with fixed $L/R_{\rm g}$. 
We found that the deviation of the spatial metric obtained by the 
cosmological Newtonian approximation from that of 
the full calculation is 
less than 1\% for $L/R_{\rm g}=20$ around the boundary of $\mathcal D$ 
and better for larger values of $L/R_{\rm g}$. 
However, the deviation of the Hubble parameter defined 
in the cosmological Newtonian approximation scheme 
and full relativistic one is 20\% 
 even for $L/R_{\rm g}=20$. 
Thus, we may say that, as expected, 
the backreaction effects of the inhomogeneities 
on the cosmic volume expansion is very large 
in the case of $L\leq20R_{\rm g}$. 
However, we may 
also say that, for 
the larger $L/R_{\rm g}$,  
the backreaction effects become smaller. 
It is worthwhile to notice that the 
backreaction effect 
acts as the brake for the cosmic 
volume expansion. 

Although our results in this paper 
agree with naive expectations, 
it is not clear by the present analysis 
if the dynamics of the black hole universe 
can be described by the EdS universe on average or not. 
Because the black hole universe cannot be exactly the EdS universe and 
the 
effect of inhomogeneities definitely exists. 
For instance, if we have an 
effective positive curvature 
term on average as the 
effect of the inhomogeneities, 
the black hole universe eventually re-collapses. 
The 
effect 
of the inhomogeneities might 
give a qualitative difference of the global expansion history 
of the universe\cite{Russ:1996km,Buchert:1999er,Nambu:2002ch,Kasai:2006bt}. 
By contrast, the present results would imply 
that the backreaction effect would not change 
the expansion law of the black hole universe from that of the EdS universe model;  
the backreaction effects might merely shift the time to the future. 
To attack this issue we need further numerical efforts and 
we leave it as a future work. 

\section*{Acknowledgements}
We thank H.~Okawa, T.~Shibata, T.~Tanaka, M.~Sasaki and N.~Yoshida 
for helpful discussions and comments. 
The numerical calculations were carried out on SR16000 at  YITP in Kyoto University.
CY is supported by a Grant-in-Aid through the
Japan Society for the Promotion of Science (JSPS).
The authors thank the Yukawa Institute for Theoretical Physics at
   Kyoto University. Discussions during the YITP workshop YITP-T-11-08 on
   ``Recent advances in numerical and analytical methods for black hole
   dynamics " were useful to complete this work.
This work was supported in part by JSPS Grant-in-Aid for Scientific
Research (C) (No. 21540276). 

\appendix
\section{Constant Mean Curvature Slices in Schwarzschild Spacetime}
\label{sec:cmc}

Let us consider the Schwarzschild spacetime, whose metric is given by 
\begin{equation}
\dd s^2=-f(r)\dd t^2+\frac{1}{f(r)}\dd r^2+r^2\dd \Omega^2, 
\end{equation}
where 
\begin{equation}
f(r)=1-\frac{\rg}{r}. 
\end{equation}
We consider a constant mean curvature(CMC) slice given by the form of 
\begin{equation}
t=h(r). 
\end{equation}
The unit normal vector is given by 
\begin{equation}
n^\mu=\frac{1}{\sqrt{f^{-1}-fh'^2}}(f^{-1},~fh',~0,~). 
\end{equation}
The CMC slice condition is given by 
\begin{eqnarray}
&&\nabla_\mu n^\mu=-K\cr
&&\Leftrightarrow\frac{1}{r^2}\del_r(r^2n^r)=-K\cr
&&\Leftrightarrow n^r=-\frac{1}{3}Kr+\frac{C}{r^2}\cr
&&\Leftrightarrow f^{-1}(1-f^2h'^2)=F(r;\rg, K, C)
:=\frac{1}{1-\frac{\rg}{r}+(-\frac{1}{3}Kr+\frac{C}{r^2})^2}, 
\end{eqnarray}
where $C$ is the integration constant. 
Then, line elements on the CMC slice is given by 
\begin{equation}
\dd\ell^2=F(r;\rg, K, C)\dd r^2+r^2\dd \Omega^2. 
\end{equation}

The transformation to the isotropic coordinate 
can be done as follows:
\begin{eqnarray}
\dd \ell^2&=&\Psi^4(\dd R^2+R^2\dd \Omega^2), \\
R&=&C\exp\left[\pm \int^r_{r_{\rm min}}\dd r \sqrt{F(r;\rg, K, C)}/r\right], \\
\Psi&=&\sqrt{r/R}, 
\end{eqnarray}
where $r_{\rm min}$ is the largest root of $F(r;\rg, K, C)=0$ and 
$r=r_{\rm min}$ corresponds to the throat. 
The minus sign is used in the region beyond 
the throat. 
We can easily check that, in the limit of $r\rightarrow \infty$, 
the isotropic coordinate $R$ has finite value if $K\neq0$. 
While $R=0$ for $r\rightarrow \infty$ if $K=0$. 
Hence, the coordinate region with $R$ 
has inside spherical boundary with $K\neq0$. 
This property is not compatible with 
the puncture method. 




\end{document}